\documentclass{icrc}

\usepackage{times}
\usepackage{graphicx} 

\begin{document}

\title{A RICH prototype for the AMS experiment}
\author[1]{G. Boudoul, T. Thuillier, A. Barrau and M. Bu\'enerd}

\affil[1]{ISN, 53 av des Martyrs, 38026 Grenoble, France}

\correspondence{buenerd@isn.in2p3.fr}

\firstpage{1}
\pubyear{2001}


\maketitle

\begin{abstract}
The AMS spectrometer will be installed on the International Space Station at
the end of 2003. Among other improvements over the first version of the
instrument, a ring imaging Cherenkov detector (RICH) will be added which latter
should open a new window for cosmic-ray physics, allowing isotope separation up
to A$\approx$25 between 1 and 10 GeV/c and elements identification up to
Z$\approx$25 between threshold and 1 TeV/c/nucleon. It should also contribute
to the high level of redundancy required for AMS and reject efficiency albedo
particles. The results of the first generation
prototype and the expected results of the new one are discussed.
\end{abstract}

\section{Introduction}

The AMS spectrometer (\cite{michel0}; \cite{Aurel} ) will be implemented on the 
International Space Station at the end of 2003 (or beginning of 2004). The instrument will 
be made of a superconducting magnet which inner volume will be mapped with a tracker 
consisting of 8 planes of silicon microstrips surrounded by a set of detectors for
particle identification : scintillator hodoscopes, electromagnetic calorimeter (ECAL), 
transition radiation detector (TRD) and ring imaging Cherenkov (RICH). This contibution 
is devoted to a study prototype aiming at the RICH definition. \\

The physics capability of the RICH counter has been investigated 
by simulations (\citet{michel1}). It should provide unique informations among the AMS 
detectors by several respects :
\begin{itemize}
\item Isotopes separation up to A$\approx$25 at best, over a momentum range extending
from about 1-2 GeV/c up to around 13 GeV/c.
\item Identification of chemical elements up to Z$\approx$25 at best, up to approximately
1 TeV/nucleon.
\item High efficiency rejection of albedo particles for momenta above the
threshold, between 1 GeV/c and 3.5 GeV/c depending on the type of
radiator (\citep{ISOLA}.
\item High level of redundancy to provide high purity samples of positrons and
antiprotons.
\item Potential help in rejecting wrongly identified antimatter candidates for
configurations similar to those found in antihelium search during phase I \citep{HELI}.
\end{itemize}
The RICH counter will allow to collect a unique sample of nuclear astrophysics
data with unprecedented statistical significance over a momentum range totally
unexplored for the most interesting isotopes.\\

\begin{figure}[t]                 
\vspace*{2.0mm} 
\includegraphics[width=8.3cm]{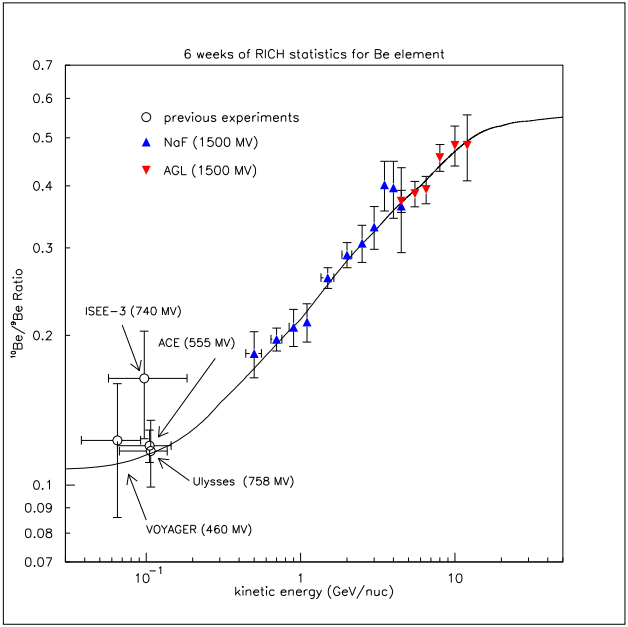} 
\caption{\label{fig:alexi}
Expected statistics for the $^{10}$Be isotope for 6 weeks of counting 
with AMS \citep{alexi}.}
\end{figure}

Fig. \ref{fig:alexi} shows, as an example, the $^{10}$Be to $^{9}$Be ratio with 6 weeks of 
counting time (\citet{alexi}). Both the number of events and the covered energy range will 
dramatically improve the available data (lower left points on the plot).\\

Recent works (\citet{david} ; \citet{fiorenza}) have emphasized the importance 
of measuring cosmic nuclei spectra for: 1) Setting strong constraints on the astrophysical 
and cosmic ray propagation parameters of the galaxy : the diffusion coefficient 
normalisation and its spectral index, the halo thickness, the Alfv\'en velocity and the 
convection velocity; 2) Increasing the sensitivity to new physics search for supersymmetric 
particles or primordial back holes; 3) Testing for the nature of the cosmic-ray sources :
 supernovae, stellar flares, Wolf-Rayet stars, etc ...

Equipped with a RICH counter, the AMS experiment will have the unique capability of being 
able to achieve the measurements of all the useful distributions with the same detector 
over the broadest range ever covered.

%

%


\section{Study prototype}

A first generation study prototype of the RICH counter has been developed and studied in 
Grenoble over the last few years (\citet{thomas}). The instrument consisted of a matrix of 
132 3/4" diameter Philips XP2802 photomultiplier tubes (PMT) available from a previous 
experiment. The size was compatible with the requirements provided by preliminary simulation 
results \citep{michel1}. The PMTs were equipped with a lime glass window (photon sensitivity 
range \{280,640\}~nm). The tubes were mounted mechanically with individual magnetic 
shieldings on a support of aluminium drilled with appropriately spaced housing holes. Each 
PMT was mounted with a socket connected by a short cable to the front end electronics board 
placed behind the matrix. The counter was installed in a vacuum chamber equipped with a 
pumping system for vacuum tests. Two experimental configurations were used for cosmic ray 
and beam particle detection respectively.\\

\begin{figure}[t]     
\vspace*{2.0mm} 
\includegraphics[width=8.3cm]{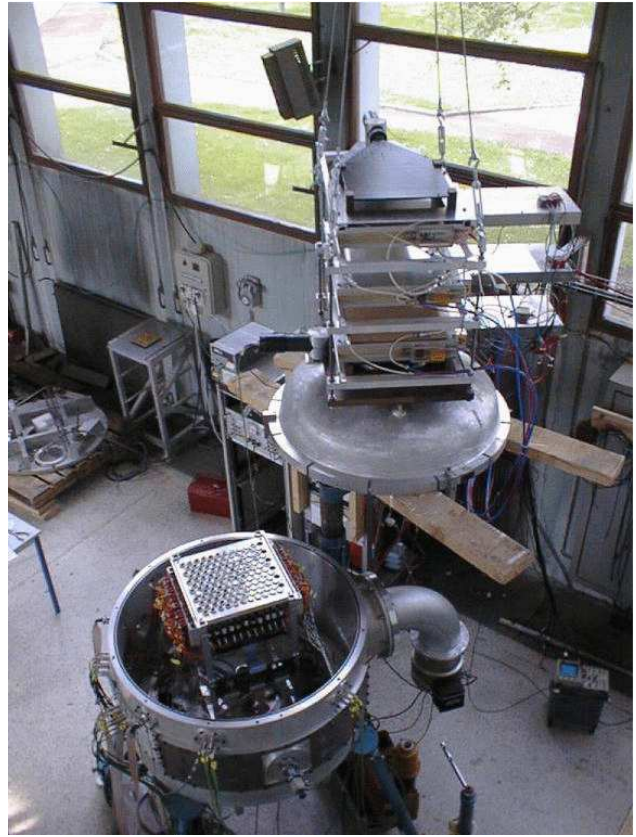} 
\caption{\label{fig:cocotte}
Photographic view of the experimental setup during cosmic ray tests, showing the PMT matrix 
placed inside the vacuum chamber, the tracker system (mwpcs) and the upper scintillator 
paddle installed on the chamber lid. The latter was moved with a crane.}
\end{figure}

In the two experimental setups used (cosmic-rays and accelerator beam), the prototype 
was complemented with a set of detectors (scintillators paddles, multiwire proportionnal 
chambers mwpcs) used to provide a trigger to the DAQ system, and to reconstruct the incident 
particle trajectory. Fig. \ref{fig:cocotte} shows a picture of the installation.
A set of radiators with refraction index from 1.025 (aerogel) up to 1.332 (sodium fluoride 
NaF) have been tested with different drift distances between radiator and detector plane, 
and different thickness. 
\begin{figure}[t]    
 \begin{center} 
 \includegraphics[scale=0.2,angle=0]{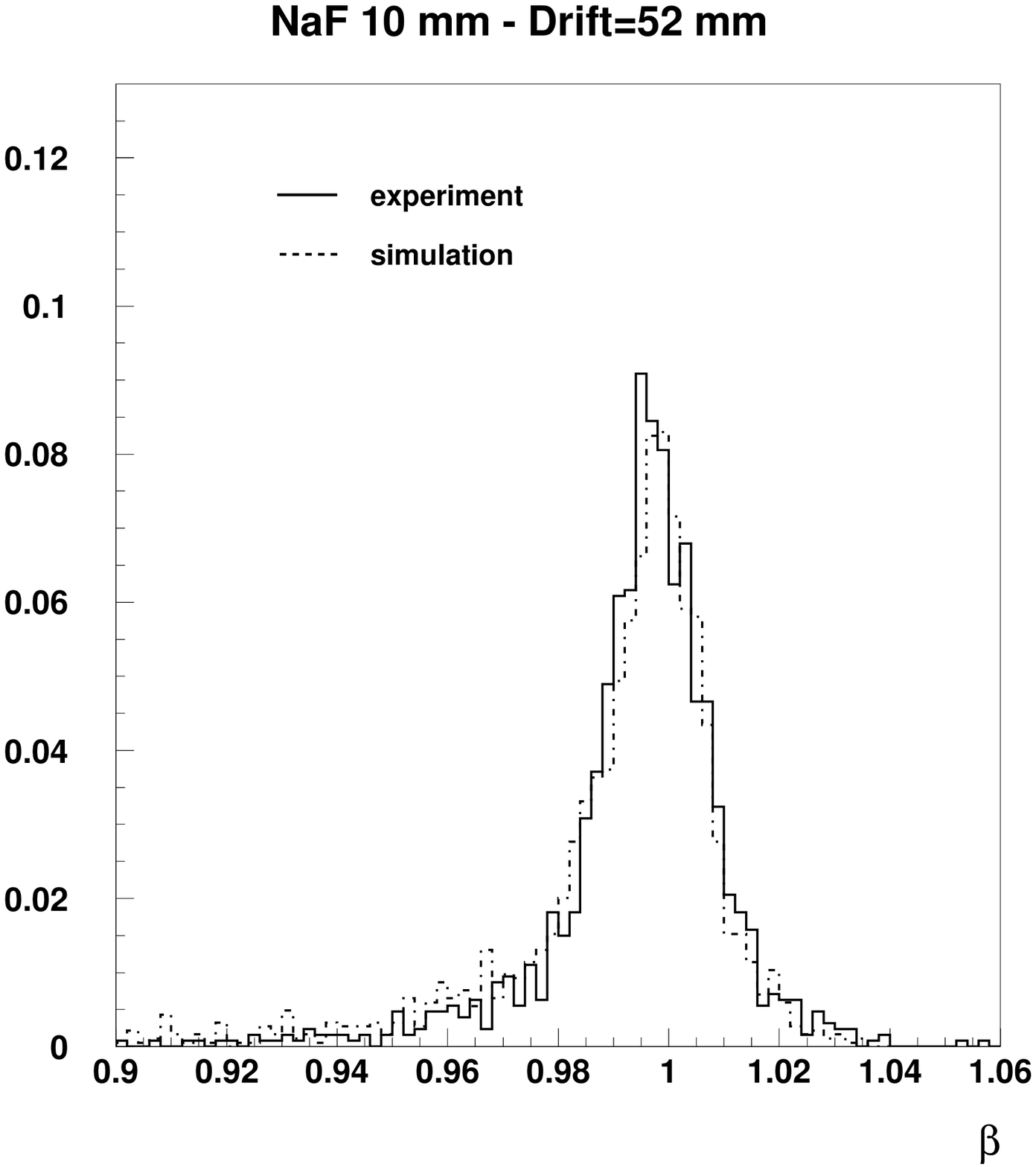}
 \includegraphics[scale=0.2,angle=0]{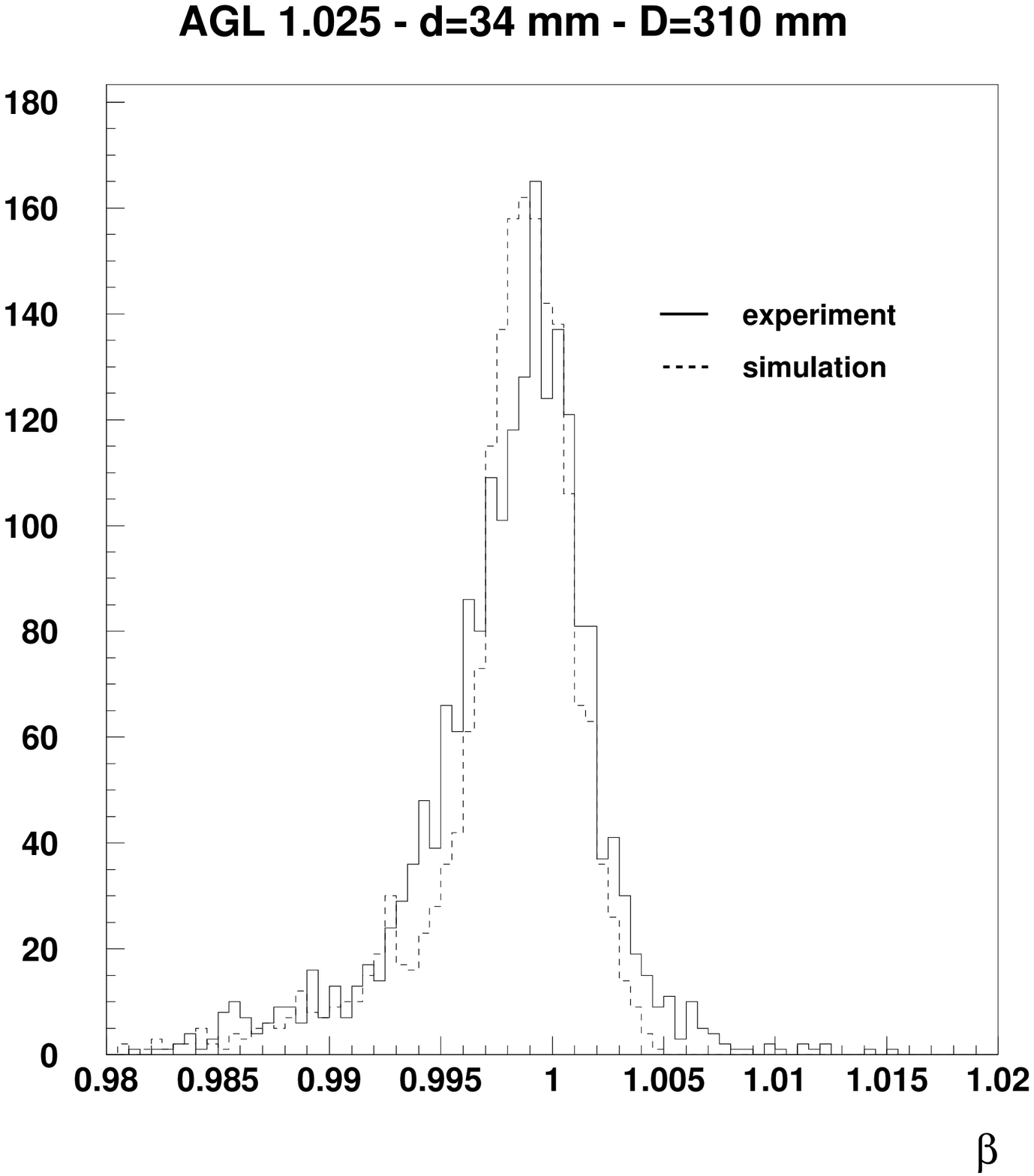}
 \end{center}
 \caption {\em Sample of cosmic test results. The experimental velocity distribution of CR
  (full line histogram) is compared with simulation results (dashed) for NaF (left) and 
aerogel (right) radiators.}
 \label{CR}
\vspace*{2.0mm} 
\includegraphics[width=8.3cm]{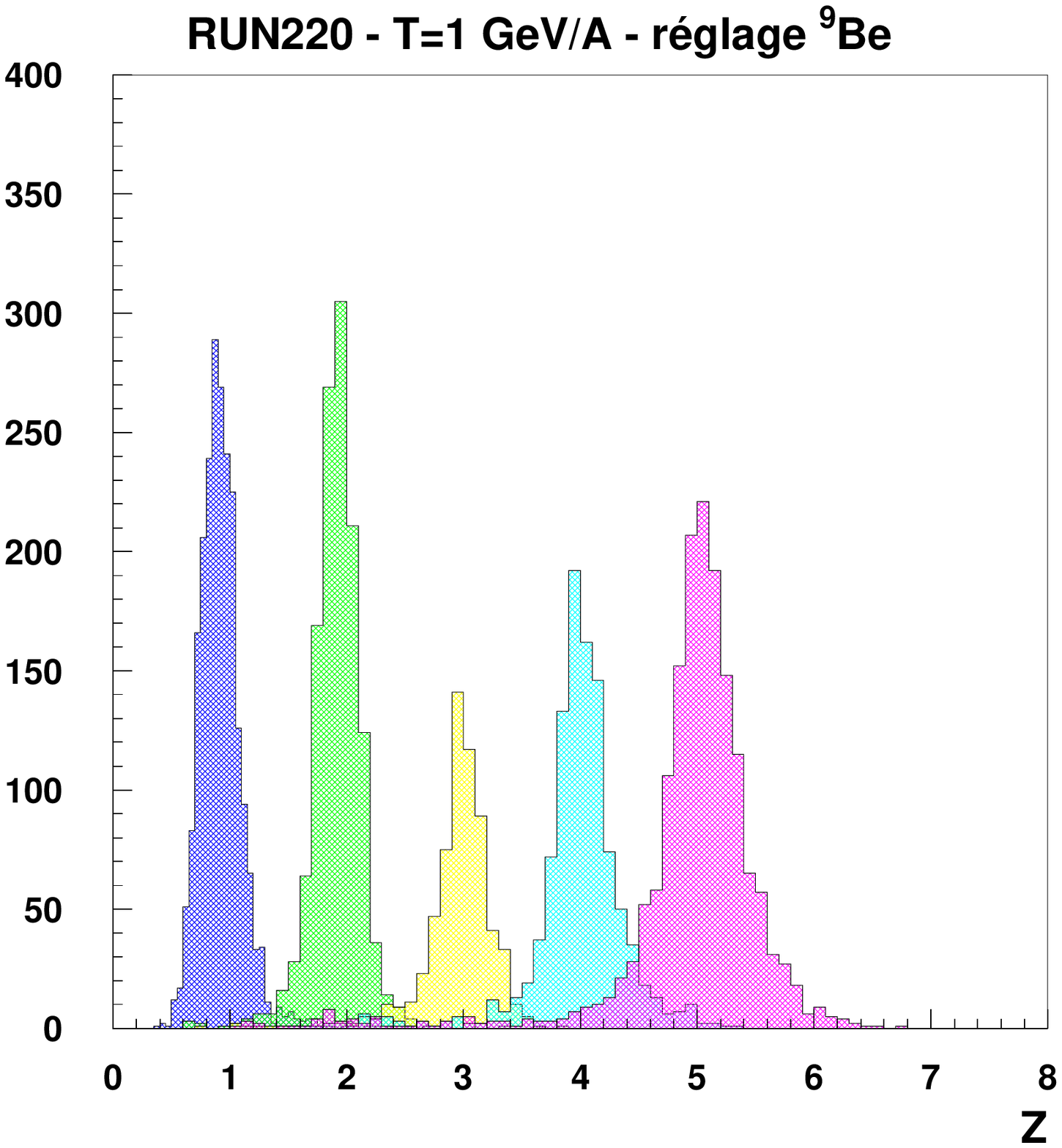} 
\caption{Experimental Z reconstruction obtained with a $^{12}$C ion beam at 
$T=1~GeV/A$. 
\label{fig:z_gsi}
}
\end{figure}
%

 
\subsection{Cosmic ray measurements}
The setup has been operated for more than a year. The results are illustrated on figure
\ref{CR}. The best experimental velocity resolution obtained for cosmic ray test was
$\frac{\delta\beta}{\beta}=0.9 \times 10^{-2}$ with $\overline{n} = 1.33$ NaF radiator 
(Cherenkov threshold around 480 MeV/nucleon), and 
$\frac{\delta\beta}{\beta}=5.2 \times 10^{-3}$ with $\overline{n} = 1.035$ aerogel radiator
(Cherenkov threshold around 3.5 GeV/nucleon). These resolutions were limited by the 
size of the photodetectors, in particular for the aerogel radiator which expected best
resolution is about 10$^{-3}$. 
The study has shown that: a) The radiator thickness and refractive index combinations for 
the final version of the counter must be optimized carefully to match Physics program;
b) This technique of proximity-focused ring imaging Cherenkov counter allows velocity 
resolutions compatible with the AMS physics purpose; and c) The good agreement between data 
and simulation gives confidence in the latter and then in the simulated performances 
expected for the final AMS RICH.
\subsection{In beam measurements}
The prototype has also been tested at the GSI / Darmstadt ion accelerator facility with 
$^{12}C$ beams of 0.6, 0.8, 1, 1.2, and 1.4 Gev/nucleon incident energies. Figure
\ref{fig:z_gsi} shows the reconstructed charge of particles for 1~GeV per nucleon beam. Beam 
particles with different masses were obtained by using a fragmentation target. Only the NaF 
radiator could be tested in this experiment since the maximum beam velocity of the 
accelerator was below the  threshold for the aerogel radiators considered.
\begin{figure}[t]       
\vspace*{2.0mm} 
\includegraphics[width=8.3cm]{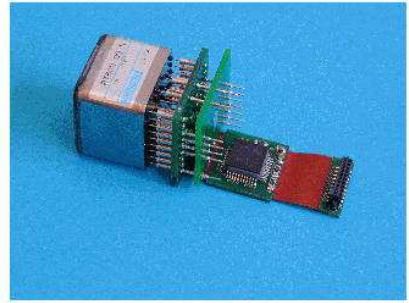} 
\caption{R5900-M16 PMT connected to the front-end electronics
\label{fig:pm}}
\end{figure}
\section{Second generation prototype}
The second generation prototype has been developed by the RICH group \footnote {AMS RICH 
group: INFN Bologna, U. Florida, ISN Grenoble, LIP Lisbon, CIEMAT Madrid, U. Maryland, and 
UNAM Mexico} of laboratories of the AMS collaboration. It will consist of one module of the 
final counter and will be operated during the second half of the year 2001.

The counter will be equipped with R7900-M16 PMTs from Hamamatsu Inc. (one hundred units in 
the prototype). The R7900-M16 is a space qualified 16 pixels PMT 
($16\,\times\,(4\times4~mm^2$)) with 
12 metal channel dynodes and a borosilicate glass window. The high voltage divider used 
is a compromise between single photoelectron resolution ($\frac{\sigma}{Q} \approx 0.5$) and 
linearity (90\% at 100 photoelectrons). The front-end electronics \citep{FE} will be placed 
next to the PMT on a flex connector linked to the readout bus (Fig. \ref{fig:pm}).
Each PMT will be equipped with solid light guides to collect the Cherenkov photons and to 
reduce the dead-space between photocathodes.
Prototype II will be installed in the same instrumental setup as the previous version and 
read with the same DAQ system. 

The main goals of this new generation of prototype are: Validation of the complex assembly 
procedure, validation of the readout electronics settings and DAQ procedure for the 16000 
output channels, investigation of the PMT+electronics response dynamics, measurement of the 
counter velocity resolution, testing the whole structure against vibrations, and validation 
of the magnetic shielding efficiency.

\section{Conclusion}
A study prototype of proximity focused RICH has been built and studied.  The results have been
found in good agreement with the simulation results. The performances should be improved 
with the next generation prototype (in account of the 3 times small pixel size) and reach 
the nominal limits of the counter. 
The latter is now under construction and should be the first step to the final counter 
assembly.

\begin{acknowledgements}
The authors are heavily indebted to the engineering team of the ISN Grenoble: J. Ballon, 
G. Barbier, R. Blanc, L. Gallin-Martel, J. Pouxe, O.Rossetto, J.P. Scordilis, and C. Vescovi
for their unvaluable help and support in the design, assembly and operation of the prototype.
\end{acknowledgements}

\balance

\end{document}